\PassOptionsToPackage{numbers}{natbib}
\documentclass{article}

\usepackage[nonatbib, final]{neurips_2020_ml4ps}

\usepackage[utf8]{inputenc} % allow utf-8 input
\usepackage[T1]{fontenc}    % use 8-bit T1 fonts
\usepackage{hyperref}       % hyperlinks
\usepackage{url}            % simple URL typesetting
\usepackage{booktabs}       % professional-quality tables
\usepackage{nicefrac}       % compact symbols for 1/2, etc.
\usepackage{microtype}      % microtypography
\usepackage{aas_macros}

\usepackage{amsmath, amsthm, amssymb, amsfonts, xspace}
\usepackage{fontawesome}
\usepackage{enumitem}
\usepackage{xspace}
\usepackage{color}
\usepackage{acronym}
\usepackage{siunitx}
\usepackage{hyperref}
\usepackage{natbib}
\usepackage{multirow}
\usepackage{graphicx}
\usepackage{comment}

\usepackage{tikz}
\usetikzlibrary{calc,
                quotes,
                positioning,
                shapes.geometric,
                decorations.markings}
                
\usepackage{pgfplots}
\pgfplotsset{compat=1.14}

\definecolor{color0}{HTML}{1F77B4}
\definecolor{color1}{HTML}{FF7F0E}
\definecolor{color2}{HTML}{2CA02C}
\definecolor{color3}{HTML}{D62728}
\definecolor{color4}{HTML}{9467BD}
\definecolor{color5}{HTML}{8C564B}
\definecolor{color6}{HTML}{E377C2}
\definecolor{color7}{HTML}{7F7F7F}
\definecolor{color8}{HTML}{BCBD22}
\definecolor{color9}{HTML}{17BECF}

\newcommand{\ie}{{i.\,e.}\xspace}

\newcommand{\stattheta}{\ensuremath{\vartheta}\xspace}

\newacro{BH}[BH]{black hole}
\newacro{GW}[GW]{gravitational wave}
\newacro{PSD}[PSD]{noise power spectral density}
\newacro{LVC}[LVC]{LIGO-Virgo Collaboration}

\newcommand{\bstattheta}{\boldsymbol \stattheta}
\newcommand{\bx}{\boldsymbol x}

\newcommand{\ULIEGE}{%
Montefiore Institute,\\
University of Li{\`e}ge, Belgium}
\newcommand{\GRAPPA}{%
Gravitation Astroparticle Physics Amsterdam (GRAPPA),\\
Institute for Theoretical Physics Amsterdam
and Delta Institute for Theoretical Physics,\\
University of Amsterdam, The Netherlands}
\newcommand{\UU}{%
Institute for Theoretical Physics,
Utrecht University, The Netherlands\\}
\newcommand{\ICG}{%
Institute for Cosmology and Gravitation,\\
University of Portsmouth, UK}

\title{Lightning-Fast Gravitational Wave Parameter Inference through Neural Amortization}

\author{%
  \AND
  Arnaud Delaunoy, Antoine Wehenkel\\
  \ULIEGE\\
  \AND
  Tanja Hinderer\\
  \UU \GRAPPA\\
  \And
  Samaya Nissanke, Christoph Weniger\\
  \GRAPPA\\
  \AND
  Andrew R. Williamson\\
  \ICG\\
  \And
  Gilles Louppe\\
  \ULIEGE\\
}

\usepackage{hyperref}

\begin{document}

\maketitle
\begin{abstract}
Gravitational waves from compact binaries measured by the LIGO and Virgo detectors are routinely analyzed using Markov Chain Monte Carlo sampling algorithms.
Because the evaluation of the likelihood function requires evaluating millions of waveform models that link between signal shapes and the source parameters, running Markov chains until convergence is typically expensive and requires days of computation. 
In this extended abstract, we provide a proof of concept that demonstrates how the latest advances in neural simulation-based inference can speed up the inference time by up to three orders of magnitude -- from days to minutes -- without impairing the performance. 
Our approach is based on a convolutional neural network modeling the likelihood-to-evidence ratio and entirely amortizes the computation of the posterior. 
We find that our model correctly estimates credible intervals for the parameters of simulated gravitational waves. 

\end{abstract}

\section{Introduction}

Inferring the parameters of sources of gravitational-waves (GW) such as compact binaries detected by LIGO~\cite{TheLIGOScientific:2014jea} and Virgo~\cite{TheVirgo:2014hva} is based on the evaluation of the Bayesian posterior probability of fifteen or more parameters that govern the shape of the signals~\cite{LIGOScientific:2019hgc}. 
Because the mapping between parameters and signals requires millions of waveform model evaluations, the analysis is computationally very expensive.  
Each likelihood evaluation requires generating the GW signal corresponding to a set of source parameters and computing its noise-weighted correlation with detector data~\cite{LIGOScientific:2019hgc}.
At present, accurate component-mass estimates only become available hours or even days after the detection of a binary black-hole coalescence.  
A full Markov-Chain Monte Carlo (MCMC) parameter estimation takes days to weeks to complete \emph{for each single event}. The improvements in sensitivity of the GW detector network for the next observing run, scheduled to start in late 2021, will lead to unprecedentedly high detection rates of binary merger events of $\sim 1$ per week to $\sim 1/$ per day~\cite{Aasi:2013wya}. 
There is thus an urgent need to develop fast and efficient methods for parameter estimation.
Furthermore, low-latency detection and characterization of GWs is essential to trigger multi-messenger time-sensitive searches in order to find electromagnetic and/or astroparticle counterparts to the GW signal~\cite{Shawhan:2019kyc}.  

Latest advances in deep learning offer an exciting direction of research towards fast detection and parameter estimation of GWs. 
For \emph{signal detection}, several previous works~\citep{George:2016hay,Li:2017chi,George:2017pmj,Gabbard:2017lja,Gebhard:2019ldz} have framed the problem as an instance of supervised classification and showcased how to make use of convolutional neural networks (CNNs) to treat GW signals.
Results presented across these work have shown fast and accurate detection both for synthetic and real data, hence confirming that CNNs are a useful and promising tool to produce real-time triggers.  
For \emph{parameter estimation}, deep learning-based alternatives to sampling algorithms have also been investigated. A large panels of neural architectures has been used, e.g. Bayesian Neural Networks~\citep{Shen:2019vep}, Conditional Variational-Autoencoders~\citep{Gabbard:2019rde}, Mixture Density Networks~\citep{Chua:2019wwt} and Normalizing Flows~\citep{green2020complete, Green:2020hst} (for a recent review see Ref.~\cite{Cuoco2020-ue}). 
These studies all point towards the fact that fast and accurate parameter estimation is within reach.

Here, we build upon previous recent developments~\cite{Cranmer:2015bka,2019arXiv190304057H,Brehmer:2019jyt} for simulation-based inference~\cite{Cranmer:2019eaq}, and demonstrate that they can drastically accelerate Bayesian posterior parameter estimation for \ac{GW} signals in realistic multi-detector scenarios.  Our approach  makes use of supervised classification models, which unlocks battle-tested neural network architectures for high-dimensional data such as CNNs,
complementary to previous works.  We find that, \textit{e.g.}, the localization of binary black-hole mergers can be accelerated by at least three orders of magnitude without compromising precision or accuracy. 

\medskip

\section{GW signal inference}
The \ac{GW} emission by the quasi-circular inspiral and merger of a binary \ac{BH} system may be characterized by eight parameters intrinsic to the system: the two \ac{BH} masses $m_{1,2}$ and the six components of the spin vectors $\boldsymbol{S}_{1,2}$.
Additionally, there are seven extrinsic parameters that relate the source position and orientation to the observer frame: the luminosity distance $d_L$, the right ascension $\alpha$ and the declination $\delta$ defining the position on the sky, the coalescence time $t_c$, the binary inclination angle $\theta_\text{JN}$, the coalescence phase angle $\Phi$ and the polarization angle $\Psi$~\cite{LIGOScientific:2019hgc}.

Here we use the waveform model
\texttt{IMRPhenomPv2}~\cite{Hannam:2013oca},
which includes a phenomenological frequency domain description of the inspiral, merger and final \ac{BH} ringdown phases.
For a proof of concept, we choose to use this model because it is very fast to evaluate while still incorporating precession effects. However the model complexity is not really a problem here because our method only requires the simulator once, to generate the training dataset.
The simulation outputs the strain observed at each detector $\mathbf{h}(t, \bstattheta)$, a vector function of the time and of the 15 waveform parameters, $\bstattheta \equiv \{ m_1, m_2, \boldsymbol{S}_1, \boldsymbol{S}_2, d_L, \alpha, \delta, t_c, \theta_\text{JN}, \Phi, \Psi \}$.  We denote synthetic and real time-series data by
$\bx = \mathbf{h} + \mathbf{n}$, where $\mathbf{n}$ refers to heteroscedastic Gaussian noise that depends on the detectors.

The task of gravitational wave parameter estimation is to determine the waveform model parameters $\bstattheta$ that are the most compatible with the strain data time series $\bx$.  
In gravitational wave analysis, this inference problem is usually framed as the computation of the Bayesian posterior distribution
\begin{equation}
    p(\bstattheta|\bx) = \frac{p(\bx|\bstattheta)p(\bstattheta)}{p(\bx)},
\end{equation}
where $p(\bx|\bstattheta)$ is the likelihood of the model parameters $\bstattheta$ given the observed data time series $\bx$, $p(\bstattheta)$ is a prior distribution over the model parameters and $p(\bx)$ is the model evidence. 

Since both the likelihood function and the prior can be evaluated, sampling algorithms such as MCMC or Nested sampling are commonly used to approximate the posterior \cite{Veitch:2014wba}. Obtaining a sufficient number of posterior samples may however be computationally expensive, taking days for binary black-hole mergers, and weeks for binary neutron-star mergers.

\bigskip

\section{Inference amortization}
We build upon previous simulation-based inference algorithms~\cite{Cranmer:2015bka,2019arXiv190304057H,Brehmer:2019jyt} to approximate the likelihood-to-evidence ratio
\begin{equation}
    r(\bx|\bstattheta) \equiv \frac{p(\bx|\bstattheta)}{p(\bx)}
\end{equation}
with a neural network. 
As demonstrated in \cite{2019arXiv190304057H}, this can be achieved by considering as positive examples (labeled $y=1$) those strain-parameter pairs $\bx,\bstattheta \sim p(\bx,\bstattheta) = p(\bx|\bstattheta) p(\bstattheta)$ which are distributed jointly,
and as negative examples (labeled $y=0$) those strain-parameter pairs $\bx,\bstattheta \sim p(\bx)p(\bstattheta)$ which are sampled marginally (independently from their respective marginal distributions).
Under mild assumptions, the decision function modeled by the Bayes optimal classifier for this binary classification problem is 
\begin{equation}
    s^*(\bx,\bstattheta) = \frac{p(\bx,\bstattheta)}{p(\bx,\bstattheta) + p(\bx)p(\bstattheta)},
\end{equation}
which recovers the likelihood-to-evidence ratio as
\begin{equation}
    \frac{s^*(\bx,\bstattheta)}{1-s^*(\bx,\bstattheta)} = \frac{p(\bx,\bstattheta)}{p(\bx)p(\bstattheta)} = \frac{p(\bx|\bstattheta)}{p(\bx)} = r(\bx|\bstattheta).
\end{equation}

After an upfront simulation and training phase, Bayesian inference for any observed strain $\bx$ is \emph{amortized}: the fast evaluation of the posterior reduces to
\begin{equation}
    \hat{p}(\bstattheta|\bx) = \hat{r}(\bx|\bstattheta)p(\bstattheta),
\end{equation}
where the evaluation of the prior $p(\bstattheta)$ is immediate while the evaluation of likelihood-to-evidence ratio $\hat{r}(\bx|\bstattheta)$ only requires a single forward pass through the neural network -- \emph{an operation which is fast and easily parallelizable in comparison to sampling algorithms.}

This amortization procedure is generic and can be applied to any Bayesian posterior inference task.  In particular, it also applies for the fast evaluation of marginal posterior distributions defined over a subset of parameters of interest, such as the posterior over the component masses $m_1-m_2$, the posterior over the mass ratio and the effective spin $q-\chi_\text{eff}$, or the posterior over right ascension and declination $\alpha-\delta$. In this case, the corresponding marginal likelihood functions are typically not available analytically and the posterior sampling procedure must be run over the entire parameter space for the only purpose of marginalizing out the other parameters. 
\medskip

\section{Network architecture and training} 
\begin{figure*}[t]
\hspace{-0.75cm}
\begin{tikzpicture}[scale=0.85, every node/.style={transform shape}]
                \tikzstyle{box} = [draw=black, fill=white, thick, rectangle, anchor=west, rounded corners, minimum height=0.7cm]
                \tikzstyle{circ} = [circle, minimum height=2.85mm, inner sep=0mm]
                \tikzstyle{cfill} = [draw=color0, very thick, fill=white]
                \tikzstyle{cdot} = [circle, fill=color0, minimum size=0.75mm, inner sep=0mm]
                \tikzstyle{dashy} = [thick, dashed, line cap=round, gray!50]
                \tikzstyle{dilation} = [anchor=center, align=center, font=\scriptsize, inner sep=0.25mm, fill=white]
                \tikzstyle{arr} = [very thick, line cap=round, postaction={decorate}, decoration={markings, mark=at position 0.55 with {\arrow{>}}}]
                \foreach \y in {3} { \foreach \x in {3.5,4.0,...,8.5} { \draw [dashy] (\x, \y-0.65) -- (\x, \y); } \foreach \x in {9.5,10.0,...,14.5} { \draw [dashy] (\x, \y-0.65) -- (\x, \y); } }
                \foreach \y in {4} { \foreach \x in {4.0,4.5,...,8.5} { \draw [dashy] (\x, \y) -- (\x, \y-1); \draw [dashy] (\x, \y) -- (\x-0.5, \y-1); } \foreach \x in {10.0,10.5,...,14.5} { \draw [dashy] (\x, \y) -- (\x, \y-1); \draw [dashy] (\x, \y) -- (\x-0.5, \y-1); } }
                \foreach \y in {5} { \foreach \x in {4.5,5.0,...,8.0} { \draw [dashy] (\x, \y) -- (\x-0.5, \y-1); \draw [dashy] (\x, \y) -- (\x+0.5, \y-1); } \foreach \x in {10.0,10.5,...,14.0} { \draw [dashy] (\x, \y) -- (\x-0.5, \y-1); \draw [dashy] (\x, \y) -- (\x+0.5, \y-1); } }
                \foreach \y in {6} { \foreach \x in {5.5,6.0,...,7.5} { \draw [dashy] (\x, \y) -- (\x-1.0, \y-1); \draw [dashy] (\x, \y) -- (\x+1.0, \y-1); } \foreach \x in {10.5,11.0,...,13.0} { \draw [dashy] (\x, \y) -- (\x-1.0, \y-1); \draw [dashy] (\x, \y) -- (\x+1.0, \y-1); } }
                
                \node [circ, fill] (1-6) at (6.5, 6) {};
                \node [circ] (1-5) at (5.5, 5) {};
                \node [circ] (2-5) at (7.5, 5) {};
                \node [circ] (1-4) at (5.0, 4) {};
                \node [circ] (2-4) at (6.0, 4) {};
                \node [circ] (3-4) at (7.0, 4) {};
                \node [circ] (4-4) at (8.0, 4) {};
                \node [circ] (1-3) at (4.5, 3) {};
                \node [circ] (2-3) at (5.0, 3) {};
                \node [circ] (3-3) at (5.5, 3) {};
                \node [circ] (4-3) at (6.0, 3) {};
                \node [circ] (5-3) at (6.5, 3) {};
                \node [circ] (6-3) at (7.0, 3) {};
                \node [circ] (7-3) at (7.5, 3) {};
                \node [circ] (8-3) at (8.0, 3) {};
                
                \draw [arr] (1-5) -- (1-6);
                \draw [arr] (2-5) -- (1-6);
                \draw [arr] (1-4) -- (1-5);
                \draw [arr] (2-4) -- (1-5);
                \draw [arr] (3-4) -- (2-5);
                \draw [arr] (4-4) -- (2-5);
                \draw [arr] (1-3) -- (1-4);
                \draw [arr] (2-3) -- (1-4);
                \draw [arr] (3-3) -- (2-4);
                \draw [arr] (4-3) -- (2-4);
                \draw [arr] (5-3) -- (3-4);
                \draw [arr] (6-3) -- (3-4);
                \draw [arr] (7-3) -- (4-4);
                \draw [arr] (8-3) -- (4-4);
                \draw [arr] (1-3) ++(0, -0.65) -- (1-3);
                \draw [arr] (2-3) ++(0, -0.65) -- (2-3);
                \draw [arr] (3-3) ++(0, -0.65) -- (3-3);
                \draw [arr] (4-3) ++(0, -0.65) -- (4-3);
                \draw [arr] (5-3) ++(0, -0.65) -- (5-3);
                \draw [arr] (6-3) ++(0, -0.65) -- (6-3);
                \draw [arr] (7-3) ++(0, -0.65) -- (7-3);
                \draw [arr] (8-3) ++(0, -0.65) -- (8-3);
                
                % DILATED CONVOLUTIONS
                \foreach \y in {3} { 
                    \foreach \x in {3.5,4.0,...,8.5} { 
                        \node [circ, cfill] at (\x, \y) {}; 
                    } 
                    \foreach \x in {9.5,10.0,...,14.5} { 
                        \node [circ, cfill] at (\x, \y) {}; 
                    } 
                }
                \foreach \y in {4} { 
                    \foreach \x in {4.0,4.5,...,8.5} { 
                        \node [circ, cfill] at (\x, \y) {}; 
                    } 
                    \foreach \x in {9.5,10.0,...,14.5} { 
                        \node [circ, cfill] at (\x, \y) {}; 
                    } 
                }
                \foreach \y in {5} { 
                    \foreach \x in {4.5,5.0,...,8.5} { 
                        \node [circ, cfill] at (\x, \y) {}; 
                    } 
                    \foreach \x in {9.5,10.0,...,14.0} { 
                        \node [circ, cfill] at (\x, \y) {}; 
                    } 
                }
                \foreach \y in {6} { 
                    \foreach \x in {5.5,6.0,...,8.5} { 
                        \node [circ, cfill] at (\x, \y) {}; 
                    } 
                    \foreach \x in {9.5,10.0,...,13.0} { 
                        \node [circ, cfill] at (\x, \y) {}; 
                    } 
                }
                
                \node [circ, fill=color0] at (6.5, 6) {};
                \foreach \y in {3, 4, 5, 6} { \foreach \x in {9.0} { \node [cdot, xshift=-1.5mm] at (\x, \y) {}; \node [cdot, xshift= 0.0mm] at (\x, \y) {}; \node [cdot, xshift= 1.5mm] at (\x, \y) {}; } }
                \foreach \y in {6.5} { \foreach \x in {5.5,6.0,...,8.5} { \node [cdot, yshift=-1.5mm] at (\x, \y) {}; \node [cdot, yshift= 0.0mm] at (\x, \y) {}; \node [cdot, yshift= 1.5mm] at (\x, \y) {}; } \foreach \x in {9.5,10.0,...,13.0} { \node [cdot, yshift=-1.5mm] at (\x, \y) {}; \node [cdot, yshift= 0.0mm] at (\x, \y) {}; \node [cdot, yshift= 1.5mm] at (\x, \y) {}; } }
                \foreach \y in {1.30} { \foreach \x in {3.5,4.0,...,14.5} { \draw [very thick, gray!50, ->] (\x, \y-0.2) -- (\x, \y+0.2); } }
                \foreach \y in {7.95} { \foreach \x in {5.5,6.0,...,13.0} { \draw [gray!50, very thick] (\x, \y-0.4) -- (\x, \y); } }
                
                %Theta line
                \draw [very thick, gray!50, -] (16.0, 0.75) -- (16.0, 7.25);
                \draw [very thick, gray!50, -] (16.0, 7.25) -- (14.5, 7.25);
                \draw [very thick, gray!50, -] (14.5, 7.25) -- (14.5, 8.25);
                
                %multilayer peceptron
                \foreach \y in {10.25} {
                \foreach \x in {5.5,6.0,...,9.5} { \draw [dashy] (\x, \y) -- (\x, \y-1); } 
                \foreach \x in {10.5, 11.0,...,14.5} { \draw [dashy] (\x, \y) -- (\x, \y-1);}
                \foreach \x in {6.0,6.5,...,9.5} {\draw [dashy] (\x, \y) -- (\x-0.5, \y-1);}
                \foreach \x in {11.0, 11.5,...,14.5} {\draw [dashy] (\x, \y) -- (\x-0.5, \y-1);}
                \foreach \x in {6.5,7.0,...,9.5} {\draw [dashy] (\x, \y) -- (\x-1.0, \y-1);}
                \foreach \x in {11.5, 12.0,...,14.5} {\draw [dashy] (\x, \y) -- (\x-1.0, \y-1);}
                \foreach \x in {5.5,6.0,...,9.0} { \draw [dashy] (\x, \y) -- (\x+0.5, \y-1); } 
                \foreach \x in {10.5, 11.0,...,14.0} { \draw [dashy] (\x, \y) -- (\x+0.5, \y-1);}
                \foreach \x in {5.5,6.0,...,8.0} { \draw [dashy] (\x, \y) -- (\x+1.0, \y-1); } 
                \foreach \x in {10.5, 11.0,...,13.5} { \draw [dashy] (\x, \y) -- (\x+1.0, \y-1);}
                }
                \draw [dashy] (9.5, 9.25) -- (10.5, 10.25);
                \draw [dashy] (9.5, 9.25) -- (11.0, 10.25);
                \draw [dashy] (9.0, 9.25) -- (10.5, 10.25);
                \draw [dashy] (9.5, 10.25) -- (10.5, 9.25);
                \draw [dashy] (9.5, 10.25) -- (11.0, 9.25);
                \draw [dashy] (9.0, 10.25) -- (10.5, 9.25);
                
                \foreach \y in {9.25} { \foreach \x in {5.5,6.0,...,9.5} { \node [circ, cfill] (\x, \y) at (\x, \y) {}; } \foreach \x in {10.5, 11.0,...,14.5} { \node [circ, cfill] at (\x, \y) {}; } }
                \foreach \y in {10.25} { \foreach \x in {5.5,6.0,...,9.5} { \node [circ, cfill] at (\x, \y) {}; } \foreach \x in {10.5, 11.0,...,14.5} { \node [circ, cfill] at (\x, \y) {}; } }
                
                \foreach \y in {9.25, 10.25} { \foreach \x in {10.0} { \node [cdot, xshift=-1.5mm] at (\x, \y) {}; \node [cdot, xshift= 0.0mm] at (\x, \y) {}; \node [cdot, xshift= 1.5mm] at (\x, \y) {}; } }
                \foreach \y in {10.75} { \foreach \x in {5.5,6.0,...,9.5} { \node [cdot, yshift=-1.5mm] at (\x, \y) {}; \node [cdot, yshift= 0.0mm] at (\x, \y) {}; \node [cdot, yshift= 1.5mm] at (\x, \y) {}; } \foreach \x in {10.5, 11.0,...,14.5} { \node [cdot, yshift=-1.5mm] at (\x, \y) {}; \node [cdot, yshift= 0.0mm] at (\x, \y) {}; \node [cdot, yshift= 1.5mm] at (\x, \y) {}; } }
                
                \foreach \y in {9.25} { \foreach \x in {5.5,6.0,...,9.5} { \draw [dashy] (\x, \y-0.65) -- (\x, \y); } \foreach \x in {10.5, 11.0,...,14.5} { \draw [dashy] (\x, \y-0.65) -- (\x, \y); } }

                \node [box, minimum width=12cm] at (3.0, 2.0) 
                    {$128 \times 8192 $};
                \node [box, minimum width=8.5cm] at (5.0, 7.25) 
                    {$128 \times 1$};
                \node [box, minimum width=1cm] at (15.5, 0.75) (label_7)
                    {$\boldsymbol{\stattheta}$};
                \node [box, minimum width=10cm] at (5.0, 8.25) (label_8)
                    {$128 + 2$};
                \node [box, minimum width=10cm] at (5.0, 11.5) 
                    {$\log r(\mathbf{x}|\mathbf{\stattheta})\in\mathbb{R}$};
                
                \node [anchor=center, align=center, inner sep=0.75mm] at (1.5, 0.5) (label_1)
                    {H1/L1 strains\\ ($2 \times 8192$)};
                \node [anchor=center, align=center, inner sep=0.75mm] at (1.5, 2.00) (label_2)
                    {Conv. layer};
                \node [anchor=center, align=center, inner sep=0.75mm] at (1.5, 4.75) (label_3)
                    {Stack of\\ 13 blocks\\ with dilated\\ Convolutional \\layers};
                \node [anchor=center, align=center, inner sep=0.75mm] at (1.5, 8.25) (label_5)
                    {Concatenation of $\boldsymbol{\stattheta}$};
                \node [anchor=center, align=center, inner sep=0.75mm] at (1.5, 9.75) (label_6)
                    {Multilayer perceptron\\ 3 layers of 200 units};
                \node [anchor=center, align=center, inner sep=0.75mm] at (1.5, 11.5) (label_9)
                    {Log likelihood-to-evidence ratio};
                \node [dilation] at (9.0, 3.50)
                    {Dilation 1};
                \node [dilation] at (9.0, 4.50)
                    {Dilation 2};
                \node [dilation] at (9.0, 5.50)
                    {Dilation 4};
                    
                \draw [very thick, ->] (label_1.north) -- (label_2.south);
                \draw [very thick, ->] (label_2.north) -- (label_3.south);
                \draw [very thick, ->] (label_3.north) -- (label_5.south);
                \draw [very thick, ->] (label_5.north) -- (label_6.south);
                \draw [very thick, ->] (label_6.north) -- (label_9.south);
                
                \draw [color3, line width=0.4mm] plot [smooth] coordinates {(3.000, 0.623) (3.020, 0.429) (3.040, 0.604) (3.060, 0.664) (3.080, 0.505) (3.100, 0.750) (3.120, 0.641) (3.140, 0.481) (3.160, 0.685) (3.180, 0.556) (3.200, 0.570) (3.220, 0.816) (3.240, 0.596) (3.260, 0.624) (3.280, 0.621) (3.300, 0.521) (3.320, 0.710) (3.340, 0.441) (3.360, 0.538) (3.380, 0.731) (3.400, 0.742) (3.420, 0.715) (3.440, 0.549) (3.460, 0.602) (3.480, 0.733) (3.500, 0.572) (3.520, 0.542) (3.540, 0.604) (3.560, 0.692) (3.580, 0.615) (3.600, 0.720) (3.620, 0.563) (3.640, 0.645) (3.660, 0.660) (3.680, 0.666) (3.700, 0.575) (3.720, 0.695) (3.740, 0.577) (3.760, 0.637) (3.780, 0.493) (3.800, 0.617) (3.820, 0.635) (3.840, 0.455) (3.860, 0.585) (3.880, 0.543) (3.900, 0.549) (3.920, 0.554) (3.940, 0.608) (3.960, 0.664) (3.980, 0.706) (4.000, 0.595) (4.020, 0.614) (4.040, 0.750) (4.060, 0.514) (4.080, 0.610) (4.100, 0.665) (4.120, 0.668) (4.140, 0.586) (4.160, 0.487) (4.180, 0.574) (4.200, 0.474) (4.220, 0.693) (4.240, 0.672) (4.260, 0.698) (4.280, 0.656) (4.300, 0.642) (4.320, 0.582) (4.340, 0.690) (4.360, 0.601) (4.380, 0.699) (4.400, 0.470) (4.420, 0.489) (4.440, 0.701) (4.460, 0.576) (4.480, 0.607) (4.500, 0.586) (4.520, 0.642) (4.540, 0.546) (4.560, 0.396) (4.580, 0.510) (4.600, 0.756) (4.620, 0.741) (4.640, 0.489) (4.660, 0.518) (4.680, 0.577) (4.700, 0.664) (4.720, 0.589) (4.740, 0.502) (4.760, 0.671) (4.780, 0.585) (4.800, 0.468) (4.820, 0.403) (4.840, 0.622) (4.860, 0.508) (4.880, 0.631) (4.900, 0.734) (4.920, 0.629) (4.940, 0.680) (4.960, 0.490) (4.980, 0.605) (5.000, 0.690) (5.020, 0.757) (5.040, 0.650) (5.060, 0.648) (5.080, 0.410) (5.100, 0.575) (5.120, 0.621) (5.140, 0.662) (5.160, 0.362) (5.180, 0.550) (5.200, 0.447) (5.220, 0.372) (5.240, 0.418) (5.260, 0.676) (5.280, 0.707) (5.300, 0.476) (5.320, 0.443) (5.340, 0.392) (5.360, 0.614) (5.380, 0.656) (5.400, 0.533) (5.420, 0.654) (5.440, 0.666) (5.460, 0.623) (5.480, 0.568) (5.500, 0.596) (5.520, 0.616) (5.540, 0.712) (5.560, 0.891) (5.580, 0.680) (5.600, 0.668) (5.620, 0.544) (5.640, 0.639) (5.660, 0.794) (5.680, 0.632) (5.700, 0.493) (5.720, 0.722) (5.740, 0.630) (5.760, 0.701) (5.780, 0.561) (5.800, 0.649) (5.820, 0.446) (5.840, 0.600) (5.860, 0.638) (5.880, 0.639) (5.900, 0.675) (5.920, 0.551) (5.940, 0.586) (5.960, 0.660) (5.980, 0.510) (6.000, 0.506) (6.020, 0.710) (6.040, 0.620) (6.060, 0.472) (6.080, 0.521) (6.100, 0.572) (6.120, 0.451) (6.140, 0.667) (6.160, 0.703) (6.180, 0.643) (6.200, 0.679) (6.220, 0.674) (6.240, 0.639) (6.260, 0.549) (6.280, 0.517) (6.300, 0.668) (6.320, 0.523) (6.340, 0.690) (6.360, 0.773) (6.380, 0.559) (6.400, 0.723) (6.420, 0.705) (6.440, 0.581) (6.460, 0.647) (6.480, 0.611) (6.500, 0.460) (6.520, 0.693) (6.540, 0.585) (6.560, 0.624) (6.580, 0.501) (6.600, 0.611) (6.620, 0.657) (6.640, 0.593) (6.660, 0.675) (6.680, 0.603) (6.700, 0.632) (6.720, 0.558) (6.740, 0.590) (6.760, 0.560) (6.780, 0.557) (6.800, 0.550) (6.820, 0.457) (6.840, 0.642) (6.860, 0.532) (6.880, 0.625) (6.900, 0.660) (6.920, 0.595) (6.940, 0.595) (6.960, 0.535) (6.980, 0.632) (7.000, 0.648) (7.020, 0.605) (7.040, 0.719) (7.060, 0.631) (7.080, 0.626) (7.100, 0.588) (7.120, 0.581) (7.140, 0.653) (7.160, 0.669) (7.180, 0.528) (7.200, 0.721) (7.220, 0.707) (7.240, 0.624) (7.260, 0.591) (7.280, 0.606) (7.300, 0.689) (7.320, 0.445) (7.340, 0.535) (7.360, 0.472) (7.380, 0.573) (7.400, 0.742) (7.420, 0.648) (7.440, 0.571) (7.460, 0.432) (7.480, 0.642) (7.500, 0.557) (7.520, 0.595) (7.540, 0.570) (7.560, 0.389) (7.580, 0.399) (7.600, 0.607) (7.620, 0.670) (7.640, 0.593) (7.660, 0.619) (7.680, 0.543) (7.700, 0.558) (7.720, 0.546) (7.740, 0.610) (7.760, 0.770) (7.780, 0.595) (7.800, 0.539) (7.820, 0.638) (7.840, 0.505) (7.860, 0.661) (7.880, 0.622) (7.900, 0.686) (7.920, 0.586) (7.940, 0.544) (7.960, 0.548) (7.980, 0.495) (8.000, 0.497) (8.020, 0.673) (8.040, 0.525) (8.060, 0.392) (8.080, 0.464) (8.100, 0.517) (8.120, 0.470) (8.140, 0.568) (8.160, 0.645) (8.180, 0.476) (8.200, 0.574) (8.220, 0.572) (8.240, 0.613) (8.260, 0.657) (8.280, 0.657) (8.300, 0.507) (8.320, 0.589) (8.340, 0.629) (8.360, 0.469) (8.380, 0.579) (8.400, 0.554) (8.420, 0.698) (8.440, 0.616) (8.460, 0.565) (8.480, 0.516) (8.500, 0.599) (8.520, 0.749) (8.540, 0.639) (8.560, 0.637) (8.580, 0.451) (8.600, 0.550) (8.620, 0.606) (8.640, 0.675) (8.660, 0.477) (8.680, 0.697) (8.700, 0.594) (8.720, 0.795) (8.740, 0.720) (8.760, 0.578) (8.780, 0.658) (8.800, 0.625) (8.820, 0.760) (8.840, 0.536) (8.860, 0.760) (8.880, 0.626) (8.900, 0.540) (8.920, 0.584) (8.940, 0.576) (8.960, 0.557) (8.980, 0.677) (9.000, 0.598) (9.020, 0.484) (9.040, 0.482) (9.060, 0.518) (9.080, 0.704) (9.100, 0.736) (9.120, 0.609) (9.140, 0.479) (9.160, 0.706) (9.180, 0.684) (9.200, 0.507) (9.220, 0.616) (9.240, 0.594) (9.260, 0.734) (9.280, 0.733) (9.300, 0.589) (9.320, 0.564) (9.340, 0.596) (9.360, 0.577) (9.380, 0.633) (9.400, 0.428) (9.420, 0.774) (9.440, 0.593) (9.460, 0.653) (9.480, 0.624) (9.500, 0.582) (9.520, 0.527) (9.540, 0.525) (9.560, 0.604) (9.580, 0.681) (9.600, 0.533) (9.620, 0.526) (9.640, 0.581) (9.660, 0.786) (9.680, 0.659) (9.700, 0.559) (9.720, 0.676) (9.740, 0.568) (9.760, 0.594) (9.780, 0.611) (9.800, 0.586) (9.820, 0.654) (9.840, 0.606) (9.860, 0.579) (9.880, 0.629) (9.900, 0.631) (9.920, 0.693) (9.940, 0.607) (9.960, 0.602) (9.980, 0.574) (10.000, 0.639) (10.020, 0.640) (10.040, 0.610) (10.060, 0.692) (10.080, 0.525) (10.100, 0.550) (10.120, 0.621) (10.140, 0.633) (10.160, 0.490) (10.180, 0.569) (10.200, 0.546) (10.220, 0.495) (10.240, 0.541) (10.260, 0.637) (10.280, 0.634) (10.300, 0.632) (10.320, 0.682) (10.340, 0.696) (10.360, 0.516) (10.380, 0.692) (10.400, 0.581) (10.420, 0.682) (10.440, 0.558) (10.460, 0.693) (10.480, 0.576) (10.500, 0.501) (10.520, 0.589) (10.540, 0.690) (10.560, 0.684) (10.580, 0.627) (10.600, 0.581) (10.620, 0.551) (10.640, 0.681) (10.660, 0.748) (10.680, 0.596) (10.700, 0.684) (10.720, 0.659) (10.740, 0.492) (10.760, 0.596) (10.780, 0.526) (10.800, 0.697) (10.820, 0.490) (10.840, 0.663) (10.860, 0.627) (10.880, 0.661) (10.900, 0.429) (10.920, 0.639) (10.940, 0.608) (10.960, 0.591) (10.980, 0.774) (11.000, 0.434) (11.020, 0.603) (11.040, 0.491) (11.060, 0.571) (11.080, 0.577) (11.100, 0.683) (11.120, 0.685) (11.140, 0.565) (11.160, 0.597) (11.180, 0.532) (11.200, 0.433) (11.220, 0.538) (11.240, 0.507) (11.260, 0.665) (11.280, 0.547) (11.300, 0.751) (11.320, 0.663) (11.340, 0.714) (11.360, 0.609) (11.380, 0.536) (11.400, 0.499) (11.420, 0.551) (11.440, 0.550) (11.460, 0.550) (11.480, 0.761) (11.500, 0.703) (11.520, 0.788) (11.540, 0.652) (11.560, 0.773) (11.580, 0.480) (11.600, 0.389) (11.620, 0.438) (11.640, 0.304) (11.660, 0.394) (11.680, 0.506) (11.700, 0.455) (11.720, 0.948) (11.740, 0.762) (11.760, 1.000) (11.780, 0.872) (11.800, 0.742) (11.820, 0.719) (11.840, 0.559) (11.860, 0.325) (11.880, 0.301) (11.900, 0.258) (11.920, 0.383) (11.940, 0.625) (11.960, 0.790) (11.980, 0.906) (12.000, 0.573) (12.020, 0.563) (12.040, 0.557) (12.060, 0.648) (12.080, 0.633) (12.100, 0.594) (12.120, 0.759) (12.140, 0.608) (12.160, 0.712) (12.180, 0.587) (12.200, 0.610) (12.220, 0.690) (12.240, 0.667) (12.260, 0.622) (12.280, 0.621) (12.300, 0.644) (12.320, 0.555) (12.340, 0.578) (12.360, 0.423) (12.380, 0.613) (12.400, 0.704) (12.420, 0.684) (12.440, 0.603) (12.460, 0.646) (12.480, 0.701) (12.500, 0.590) (12.520, 0.572) (12.540, 0.757) (12.560, 0.639) (12.580, 0.632) (12.600, 0.551) (12.620, 0.599) (12.640, 0.601) (12.660, 0.343) (12.680, 0.530) (12.700, 0.541) (12.720, 0.599) (12.740, 0.626) (12.760, 0.518) (12.780, 0.599) (12.800, 0.620) (12.820, 0.670) (12.840, 0.669) (12.860, 0.520) (12.880, 0.533) (12.900, 0.512) (12.920, 0.517) (12.940, 0.633) (12.960, 0.628) (12.980, 0.588) (13.000, 0.472) (13.020, 0.558) (13.040, 0.613) (13.060, 0.608) (13.080, 0.636) (13.100, 0.421) (13.120, 0.571) (13.140, 0.623) (13.160, 0.396) (13.180, 0.606) (13.200, 0.717) (13.220, 0.813) (13.240, 0.687) (13.260, 0.653) (13.280, 0.622) (13.300, 0.508) (13.320, 0.527) (13.340, 0.536) (13.360, 0.560) (13.380, 0.488) (13.400, 0.522) (13.420, 0.493) (13.440, 0.528) (13.460, 0.576) (13.480, 0.638) (13.500, 0.616) (13.520, 0.563) (13.540, 0.586) (13.560, 0.558) (13.580, 0.599) (13.600, 0.663) (13.620, 0.692) (13.640, 0.543) (13.660, 0.510) (13.680, 0.500) (13.700, 0.511) (13.720, 0.548) (13.740, 0.499) (13.760, 0.657) (13.780, 0.617) (13.800, 0.613) (13.820, 0.656) (13.840, 0.546) (13.860, 0.592) (13.880, 0.669) (13.900, 0.552) (13.920, 0.494) (13.940, 0.449) (13.960, 0.653) (13.980, 0.667) (14.000, 0.576) (14.020, 0.445) (14.040, 0.509) (14.060, 0.816) (14.080, 0.608) (14.100, 0.570) (14.120, 0.615) (14.140, 0.654) (14.160, 0.673) (14.180, 0.361) (14.200, 0.643) (14.220, 0.648) (14.240, 0.611) (14.260, 0.595) (14.280, 0.565) (14.300, 0.572) (14.320, 0.607) (14.340, 0.625) (14.360, 0.559) (14.380, 0.717) (14.400, 0.676) (14.420, 0.623) (14.440, 0.733) (14.460, 0.621) (14.480, 0.593) (14.500, 0.540) (14.520, 0.699) (14.540, 0.491) (14.560, 0.614) (14.580, 0.591) (14.600, 0.644) (14.620, 0.633) (14.640, 0.565) (14.660, 0.676) (14.680, 0.582) (14.700, 0.726) (14.720, 0.628) (14.740, 0.704) (14.760, 0.545) (14.780, 0.558) (14.800, 0.528) (14.820, 0.495) (14.840, 0.709) (14.860, 0.581) (14.880, 0.854) (14.900, 0.569) (14.920, 0.617) (14.940, 0.596) (14.960, 0.572) (14.980, 0.660) };
                
                \draw [color1, line width=0.4mm] plot [smooth] coordinates{ (3.000, +0.773) (3.020, +0.579) (3.040, +0.754) (3.060, +0.814) (3.080, +0.655) (3.100, +0.900) (3.120, +0.791) (3.140, +0.631) (3.160, +0.835) (3.180, +0.706) (3.200, +0.720) (3.220, +0.966) (3.240, +0.746) (3.260, +0.774) (3.280, +0.771) (3.300, +0.671) (3.320, +0.860) (3.340, +0.591) (3.360, +0.688) (3.380, +0.881) (3.400, +0.892) (3.420, +0.865) (3.440, +0.699) (3.460, +0.752) (3.480, +0.883) (3.500, +0.722) (3.520, +0.692) (3.540, +0.754) (3.560, +0.842) (3.580, +0.765) (3.600, +0.870) (3.620, +0.713) (3.640, +0.795) (3.660, +0.810) (3.680, +0.816) (3.700, +0.725) (3.720, +0.845) (3.740, +0.727) (3.760, +0.787) (3.780, +0.643) (3.800, +0.767) (3.820, +0.785) (3.840, +0.605) (3.860, +0.735) (3.880, +0.693) (3.900, +0.699) (3.920, +0.704) (3.940, +0.758) (3.960, +0.814) (3.980, +0.856) (4.000, +0.745) (4.020, +0.764) (4.040, +0.900) (4.060, +0.664) (4.080, +0.760) (4.100, +0.815) (4.120, +0.818) (4.140, +0.736) (4.160, +0.637) (4.180, +0.724) (4.200, +0.624) (4.220, +0.843) (4.240, +0.822) (4.260, +0.848) (4.280, +0.806) (4.300, +0.792) (4.320, +0.732) (4.340, +0.840) (4.360, +0.751) (4.380, +0.849) (4.400, +0.620) (4.420, +0.639) (4.440, +0.851) (4.460, +0.726) (4.480, +0.757) (4.500, +0.736) (4.520, +0.792) (4.540, +0.696) (4.560, +0.546) (4.580, +0.660) (4.600, +0.906) (4.620, +0.891) (4.640, +0.639) (4.660, +0.668) (4.680, +0.727) (4.700, +0.814) (4.720, +0.739) (4.740, +0.652) (4.760, +0.821) (4.780, +0.735) (4.800, +0.618) (4.820, +0.553) (4.840, +0.772) (4.860, +0.658) (4.880, +0.781) (4.900, +0.884) (4.920, +0.779) (4.940, +0.830) (4.960, +0.640) (4.980, +0.755) (5.000, +0.840) (5.020, +0.907) (5.040, +0.800) (5.060, +0.798) (5.080, +0.560) (5.100, +0.725) (5.120, +0.771) (5.140, +0.812) (5.160, +0.512) (5.180, +0.700) (5.200, +0.597) (5.220, +0.522) (5.240, +0.568) (5.260, +0.826) (5.280, +0.857) (5.300, +0.626) (5.320, +0.593) (5.340, +0.542) (5.360, +0.764) (5.380, +0.806) (5.400, +0.683) (5.420, +0.804) (5.440, +0.816) (5.460, +0.773) (5.480, +0.718) (5.500, +0.746) (5.520, +0.766) (5.540, +0.862) (5.560, +1.041) (5.580, +0.830) (5.600, +0.818) (5.620, +0.694) (5.640, +0.789) (5.660, +0.944) (5.680, +0.782) (5.700, +0.643) (5.720, +0.872) (5.740, +0.780) (5.760, +0.851) (5.780, +0.711) (5.800, +0.799) (5.820, +0.596) (5.840, +0.750) (5.860, +0.788) (5.880, +0.789) (5.900, +0.825) (5.920, +0.701) (5.940, +0.736) (5.960, +0.810) (5.980, +0.660) (6.000, +0.656) (6.020, +0.860) (6.040, +0.770) (6.060, +0.622) (6.080, +0.671) (6.100, +0.722) (6.120, +0.601) (6.140, +0.817) (6.160, +0.853) (6.180, +0.793) (6.200, +0.829) (6.220, +0.824) (6.240, +0.789) (6.260, +0.699) (6.280, +0.667) (6.300, +0.818) (6.320, +0.673) (6.340, +0.840) (6.360, +0.923) (6.380, +0.709) (6.400, +0.873) (6.420, +0.855) (6.440, +0.731) (6.460, +0.797) (6.480, +0.761) (6.500, +0.610) (6.520, +0.843) (6.540, +0.735) (6.560, +0.774) (6.580, +0.651) (6.600, +0.761) (6.620, +0.807) (6.640, +0.743) (6.660, +0.825) (6.680, +0.753) (6.700, +0.782) (6.720, +0.708) (6.740, +0.740) (6.760, +0.710) (6.780, +0.707) (6.800, +0.700) (6.820, +0.607) (6.840, +0.792) (6.860, +0.682) (6.880, +0.775) (6.900, +0.810) (6.920, +0.745) (6.940, +0.745) (6.960, +0.685) (6.980, +0.782) (7.000, +0.798) (7.020, +0.755) (7.040, +0.869) (7.060, +0.781) (7.080, +0.776) (7.100, +0.738) (7.120, +0.731) (7.140, +0.803) (7.160, +0.819) (7.180, +0.678) (7.200, +0.871) (7.220, +0.857) (7.240, +0.774) (7.260, +0.741) (7.280, +0.756) (7.300, +0.839) (7.320, +0.595) (7.340, +0.685) (7.360, +0.622) (7.380, +0.723) (7.400, +0.892) (7.420, +0.798) (7.440, +0.721) (7.460, +0.582) (7.480, +0.792) (7.500, +0.707) (7.520, +0.745) (7.540, +0.720) (7.560, +0.539) (7.580, +0.549) (7.600, +0.757) (7.620, +0.820) (7.640, +0.743) (7.660, +0.769) (7.680, +0.693) (7.700, +0.708) (7.720, +0.696) (7.740, +0.760) (7.760, +0.920) (7.780, +0.745) (7.800, +0.689) (7.820, +0.788) (7.840, +0.655) (7.860, +0.811) (7.880, +0.772) (7.900, +0.836) (7.920, +0.736) (7.940, +0.694) (7.960, +0.698) (7.980, +0.645) (8.000, +0.647) (8.020, +0.823) (8.040, +0.675) (8.060, +0.542) (8.080, +0.614) (8.100, +0.667) (8.120, +0.620) (8.140, +0.718) (8.160, +0.795) (8.180, +0.626) (8.200, +0.724) (8.220, +0.722) (8.240, +0.763) (8.260, +0.807) (8.280, +0.807) (8.300, +0.657) (8.320, +0.739) (8.340, +0.779) (8.360, +0.619) (8.380, +0.729) (8.400, +0.704) (8.420, +0.848) (8.440, +0.766) (8.460, +0.715) (8.480, +0.666) (8.500, +0.749) (8.520, +0.899) (8.540, +0.789) (8.560, +0.787) (8.580, +0.601) (8.600, +0.700) (8.620, +0.756) (8.640, +0.825) (8.660, +0.627) (8.680, +0.847) (8.700, +0.744) (8.720, +0.945) (8.740, +0.870) (8.760, +0.728) (8.780, +0.808) (8.800, +0.775) (8.820, +0.910) (8.840, +0.686) (8.860, +0.910) (8.880, +0.776) (8.900, +0.690) (8.920, +0.734) (8.940, +0.726) (8.960, +0.707) (8.980, +0.827) (9.000, +0.748) (9.020, +0.634) (9.040, +0.632) (9.060, +0.668) (9.080, +0.854) (9.100, +0.886) (9.120, +0.759) (9.140, +0.629) (9.160, +0.856) (9.180, +0.834) (9.200, +0.657) (9.220, +0.766) (9.240, +0.744) (9.260, +0.884) (9.280, +0.883) (9.300, +0.739) (9.320, +0.714) (9.340, +0.746) (9.360, +0.727) (9.380, +0.783) (9.400, +0.578) (9.420, +0.924) (9.440, +0.743) (9.460, +0.803) (9.480, +0.774) (9.500, +0.732) (9.520, +0.677) (9.540, +0.675) (9.560, +0.754) (9.580, +0.831) (9.600, +0.683) (9.620, +0.676) (9.640, +0.731) (9.660, +0.936) (9.680, +0.809) (9.700, +0.709) (9.720, +0.826) (9.740, +0.718) (9.760, +0.744) (9.780, +0.761) (9.800, +0.736) (9.820, +0.804) (9.840, +0.756) (9.860, +0.729) (9.880, +0.779) (9.900, +0.781) (9.920, +0.843) (9.940, +0.757) (9.960, +0.752) (9.980, +0.724) (10.000, +0.789) (10.020, +0.790) (10.040, +0.760) (10.060, +0.842) (10.080, +0.675) (10.100, +0.700) (10.120, +0.771) (10.140, +0.783) (10.160, +0.640) (10.180, +0.719) (10.200, +0.696) (10.220, +0.645) (10.240, +0.691) (10.260, +0.787) (10.280, +0.784) (10.300, +0.782) (10.320, +0.832) (10.340, +0.846) (10.360, +0.666) (10.380, +0.842) (10.400, +0.731) (10.420, +0.832) (10.440, +0.708) (10.460, +0.843) (10.480, +0.726) (10.500, +0.651) (10.520, +0.739) (10.540, +0.840) (10.560, +0.834) (10.580, +0.777) (10.600, +0.731) (10.620, +0.701) (10.640, +0.831) (10.660, +0.898) (10.680, +0.746) (10.700, +0.834) (10.720, +0.809) (10.740, +0.642) (10.760, +0.746) (10.780, +0.676) (10.800, +0.847) (10.820, +0.640) (10.840, +0.813) (10.860, +0.777) (10.880, +0.811) (10.900, +0.579) (10.920, +0.789) (10.940, +0.758) (10.960, +0.741) (10.980, +0.924) (11.000, +0.584) (11.020, +0.753) (11.040, +0.641) (11.060, +0.721) (11.080, +0.727) (11.100, +0.833) (11.120, +0.835) (11.140, +0.715) (11.160, +0.747) (11.180, +0.682) (11.200, +0.583) (11.220, +0.688) (11.240, +0.657) (11.260, +0.815) (11.280, +0.697) (11.300, +0.901) (11.320, +0.813) (11.340, +0.864) (11.360, +0.759) (11.380, +0.686) (11.400, +0.649) (11.420, +0.701) (11.440, +0.700) (11.460, +0.700) (11.480, +0.911) (11.500, +0.853) (11.520, +0.938) (11.540, +0.802) (11.560, +0.923) (11.580, +0.630) (11.600, +0.539) (11.620, +0.588) (11.640, +0.454) (11.660, +0.544) (11.680, +0.656) (11.700, +0.605) (11.720, +1.098) (11.740, +0.912) (11.760, +1.150) (11.780, +1.022) (11.800, +0.892) (11.820, +0.869) (11.840, +0.709) (11.860, +0.475) (11.880, +0.451) (11.900, +0.408) (11.920, +0.533) (11.940, +0.775) (11.960, +0.940) (11.980, +1.056) (12.000, +0.723) (12.020, +0.713) (12.040, +0.707) (12.060, +0.798) (12.080, +0.783) (12.100, +0.744) (12.120, +0.909) (12.140, +0.758) (12.160, +0.862) (12.180, +0.737) (12.200, +0.760) (12.220, +0.840) (12.240, +0.817) (12.260, +0.772) (12.280, +0.771) (12.300, +0.794) (12.320, +0.705) (12.340, +0.728) (12.360, +0.573) (12.380, +0.763) (12.400, +0.854) (12.420, +0.834) (12.440, +0.753) (12.460, +0.796) (12.480, +0.851) (12.500, +0.740) (12.520, +0.722) (12.540, +0.907) (12.560, +0.789) (12.580, +0.782) (12.600, +0.701) (12.620, +0.749) (12.640, +0.751) (12.660, +0.493) (12.680, +0.680) (12.700, +0.691) (12.720, +0.749) (12.740, +0.776) (12.760, +0.668) (12.780, +0.749) (12.800, +0.770) (12.820, +0.820) (12.840, +0.819) (12.860, +0.670) (12.880, +0.683) (12.900, +0.662) (12.920, +0.667) (12.940, +0.783) (12.960, +0.778) (12.980, +0.738) (13.000, +0.622) (13.020, +0.708) (13.040, +0.763) (13.060, +0.758) (13.080, +0.786) (13.100, +0.571) (13.120, +0.721) (13.140, +0.773) (13.160, +0.546) (13.180, +0.756) (13.200, +0.867) (13.220, +0.963) (13.240, +0.837) (13.260, +0.803) (13.280, +0.772) (13.300, +0.658) (13.320, +0.677) (13.340, +0.686) (13.360, +0.710) (13.380, +0.638) (13.400, +0.672) (13.420, +0.643) (13.440, +0.678) (13.460, +0.726) (13.480, +0.788) (13.500, +0.766) (13.520, +0.713) (13.540, +0.736) (13.560, +0.708) (13.580, +0.749) (13.600, +0.813) (13.620, +0.842) (13.640, +0.693) (13.660, +0.660) (13.680, +0.650) (13.700, +0.661) (13.720, +0.698) (13.740, +0.649) (13.760, +0.807) (13.780, +0.767) (13.800, +0.763) (13.820, +0.806) (13.840, +0.696) (13.860, +0.742) (13.880, +0.819) (13.900, +0.702) (13.920, +0.644) (13.940, +0.599) (13.960, +0.803) (13.980, +0.817) (14.000, +0.726) (14.020, +0.595) (14.040, +0.659) (14.060, +0.966) (14.080, +0.758) (14.100, +0.720) (14.120, +0.765) (14.140, +0.804) (14.160, +0.823) (14.180, +0.511) (14.200, +0.793) (14.220, +0.798) (14.240, +0.761) (14.260, +0.745) (14.280, +0.715) (14.300, +0.722) (14.320, +0.757) (14.340, +0.775) (14.360, +0.709) (14.380, +0.867) (14.400, +0.826) (14.420, +0.773) (14.440, +0.883) (14.460, +0.771) (14.480, +0.743) (14.500, +0.690) (14.520, +0.849) (14.540, +0.641) (14.560, +0.764) (14.580, +0.741) (14.600, +0.794) (14.620, +0.783) (14.640, +0.715) (14.660, +0.826) (14.680, +0.732) (14.700, +0.876) (14.720, +0.778) (14.740, +0.854) (14.760, +0.695) (14.780, +0.708) (14.800, +0.678) (14.820, +0.645) (14.840, +0.859) (14.860, +0.731) (14.880, +1.004) (14.900, +0.719) (14.920, +0.767) (14.940, +0.746) (14.960, +0.722) (14.980, +0.810) };
                
\end{tikzpicture}
\caption{The architecture of the neural ratio estimator, adapted from \cite{Gebhard:2019ldz}. A feature map of both the Hanford's and Livingston's strains is produced by a CNN composed of a 1D convolutional layer ($128$ kernels of size $1$) followed by $13$ dilated convolutional layers ($128$ kernels of size $2$). Finally, the convolutional feature map ($128$ channels of size $1$) is concatenated with the parameters $\bstattheta$ and fed to a 3-layer fully connected network for approximating the log likelihood-to-evidence ratio.}
\label{fig:nn}
\end{figure*}
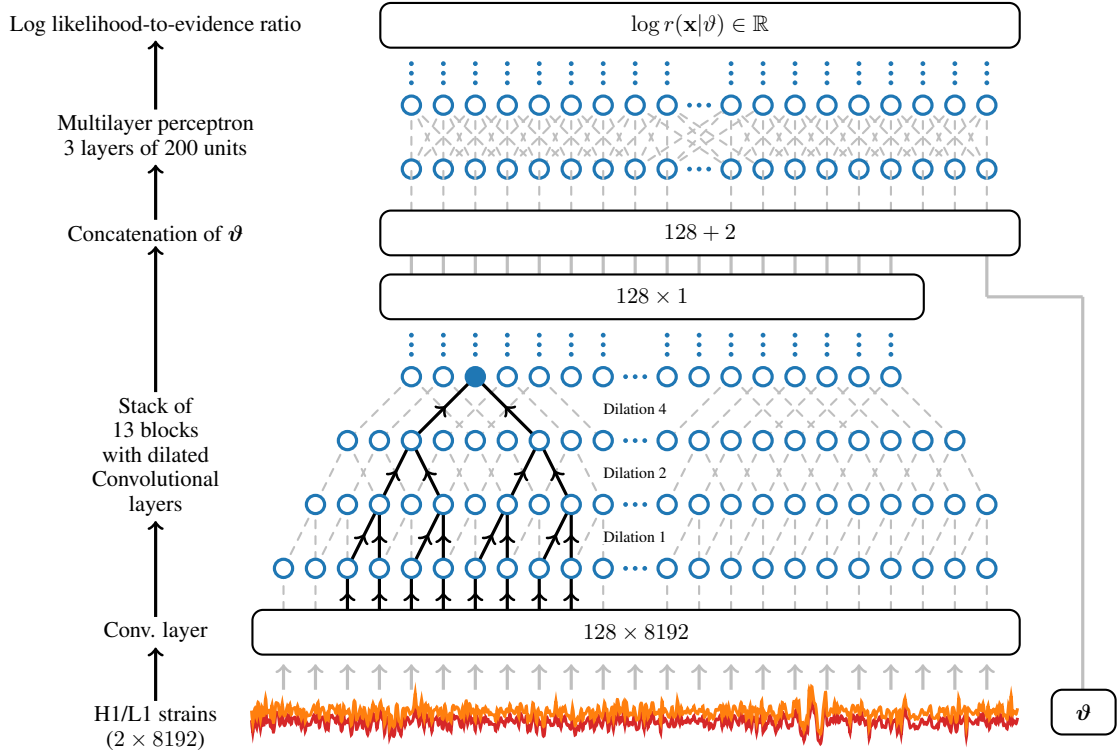

The neural network used to model the likelihood-to-evidence ratio and to demonstrate our inference method on GW signals is illustrated in Figure~\ref{fig:nn}. The Hanford and Livingston detectors input strains $\bx$ are processed by a convolutional trunk made of 13 layers of dilated convolutions. The output log-ratio is  computed by a 3-layer  fully connected network ending with no activation function and taking as input both the convolutional feature map and the parameters $\bstattheta$. At training time, the class $y=1$ probability $p(y=1|\bx, \bstattheta)$ is computed as the sigmoid activation $\sigma(\log r(\bx|\bstattheta))$ of the network's output.

To train and evaluate our neural network model, we consider simulated signals of binary black hole mergers using \texttt{IMRPhenomPv2} as approximant (including the effects of spin precession), and parameters drawn from the prior distribution $p(\bstattheta)$ summarized in Table~\ref{tab:prior}. To ease comparison with standard analysis results, the prior is chosen identical to the analysis of GW150914~\cite{TheLIGOScientific:2016wfe} and the companion analysis of $\textsc{PyCBC}$~\cite{Biwer:2018osg}. 
Our data generation process follows the procedure of \cite{Gebhard:2019ldz} in order to produce realistic synthetic training data. In short, background noise is emulated using a Gaussian noise model with a PSD identical to the one used in \cite{abbott2019gwtc}. Simulated signals are injected into the noise and the result is then whitened and high-passed at 20 Hz to remove simulation artefacts. Finally, the generated strains are cropped to a length of $4s$ in such a way that the maximum of the signal always ends up at a random location in a fixed interval of $0.2s$ within the sample. Using a sampling rate of 2048 Hz, the resulting observable $\bx$ for each realization is an array of size $8192 \times 2$, where the channels correspond to the strains from each of the two interferometers. Using this procedure, we generated $10^6$ observables for training, $2\times 10^5$ for validation and $2\times 10^5$ for test. 

\section{Results}

To evaluate the performance of our method, we perform inference on simulated gravitational waves. Figure \ref{fig:posteriors} showcases test GWs and the resulting approximate posteriors. The first row compares the credible intervals obtained with our method and with MCMC on a signal similar to GW150914. Both MCMC and our method produce consistent results but \textit{our method takes less than a minute to run while MCMC ran for more than a day!} Since performing an MCMC run is computationally expensive, the following rows only compare against the nominal parameter values that were used to generate the signals. We observe that our method is often able to infer the parameters with high precision and to produce narrow credible intervals. Sometimes, our method produces wide intervals.

We look at the coverage on 1000 simulated signals to evaluate the reliability of the estimated credible intervals. 
If the model estimates correctly the credible intervals, then the empirical coverage probability should be close to the nominal coverage probability.
For instance, the estimated 50\%-credible intervals for pairs $\bx, \bstattheta \sim p(\bx,\bstattheta)$ should contain the nominal value $\bstattheta$ approximately 50\% of the time. 
As reported in Table~\ref{tab:coverage}, the empirical coverage probability is usually slightly higher than expected. This shows that the model is slightly under-confident in its predictions and hence produces credible intervals than are slightly larger than expected. This under-confidence of the model is however moderated and does not preclude the contours to be useful in practice since predictions are conservative.

\begin{table}
    \centering
    \begin{tabular}{|c|c|c|c|c|}
        \hline
         & $m_{max}^{det}$, $m_{min}^{det}$ & $d_L$,$\theta_{JN}$ & $\chi_{\text{eff}}$, $q$ & $\alpha, \delta$ \\
        \hline
        50\% credible interval coverage & 51.8\% & 50.7\% & 57.5\% & 54.2\%\\
        \hline
        90\% credible interval coverage & 90.2\% & 89.8\% & 93.3\% & 92.8\%\\
        \hline
    \end{tabular}
    \label{tab:coverage}\\
    \vspace{1em}
    \caption{Empirical coverage probability of the estimated credible intervals.}
\end{table}

\section{Conclusions}

In this paper, we provide a proof of concept that demonstrates that neural simulation-based inference, and more specifically likelihood-to-evidence ratio estimation,  can be used to speed up the analysis of GWs by up to three orders of magnitude while producing good posterior parameter distributions. Although these preliminary results are promising, we note that there are still obstacles to overcome before deploying neural simulation-based inference in official GW analysis pipelines.
Possible improvements include accounting for the measured noise PSD, as well as further exploring the preprocessing of the data.
Further assessments of the statistical validity of the estimated posteriors would also be needed before making any reliable scientific claims.
Recently, complementary deep learning approaches such as normalizing flows and conditional variational auto-encoders have also been applied to the same problem. A detailed comparison between all those methods is required to further advance the field.

\section*{Broader impact}

Neural simulation-based inference has the potential to speed up scientific discoveries by allowing the fast analysis of gravitational waves. It also unlocks multi-messenger astronomy as the fast inference of the binary black-holes merger's sky position is required to measure electromagnetic and/or astroparticle counterparts to the GW signal. Malicious usage of our research is hard to imagine. However, failure modes of our inference engine could lead to erroneous scientific claims, and thus should be carefully validated.

\section*{Acknowledgments}
AW is a research fellow of the F.R.S.-FNRS (Belgium) and acknowledges its financial support. GL is recipient of the ULiege - NRB Chair on Big data and is thankful for the support of NRB. TH acknowledges support from NWO Projectruimte grant GWEM-NS and the DeltaITP. ARW is supported by STFC grant ST/S000550/1. SN is grateful for support from NWO VIDI and Projectruimte Grants of the Innovational Research Incentives Scheme (Vernieuwingsimpuls) financed by the Netherlands Organization for Scientific Research (NWO). We are grateful for computational resources provided by Cardiff University, and funded by an STFC grant supporting UK Involvement in the Operation of Advanced LIGO.

\appendix
\bibliography{gw-lfi}

\newpage

\appendix

\section{Likelihood function and prior distribution}

Our likelihood function (which is only defined up to an overall constant factor) is given by
\begin{equation}
    \label{eq:likelihood}
    \begin{split}
        &p(d(t)|\bstattheta, h(t,\bstattheta)) = {}\\
        &\exp \left\{ -\frac{1}{2} \sum_{k=1}^N \left\langle \tilde{d}_k(f) - \tilde{h}_k(f,\bstattheta) \middle| \tilde{d}_k(f) - \tilde{h}_k(f,\bstattheta) \right\rangle \right\} \, ,
    \end{split}
\end{equation}
where $\langle a | b \rangle$ denotes the noise-weighted inner product, defined as
\begin{equation}
    \left\langle \tilde{a}_k(f) \middle| \tilde{b}_k(f) \right\rangle = 4 \Re \int_0^\infty \frac{\tilde{a}_k(f) \tilde{b}_k(f)}{S_k(f)} \, df \, ,
\end{equation}

$d(t)$ are the \ac{GW} detector data time series, $N$ is the number of detectors, $\tilde{d}_k(f)$ is the Fourier transform of the data from the $k$-th detector, $S_k(f)$ is the \ac{PSD} of the $k$-th detector, and $\tilde{h}_k(f,\bstattheta)$ is the frequency domain model waveform projected onto the $k$-th detector. Specifically, this is the sum of the $+$ and $\times$ \ac{GW} polarizations weighted by the antenna response factors $F^{+,\times}$, which depend on the extrinsic parameters relating the source position and orientation to the detectors, \ie
\begin{equation}
    \tilde{h}_k(f,\bstattheta) = F^+_k \tilde{h}^+(f,\bstattheta) + F^\times_k \tilde{h}^\times(f,\bstattheta).
\end{equation}

Table~\ref{tab:prior} summarizes the signal model parameters and their respective priors, as used in the training of the inference network. We adopt the same prior as in  \cite{TheLIGOScientific:2016wfe, Biwer:2018osg}, to which we refer for further details.

\begin{table*}[h]
    \centering 
    \begin{tabular}{l c c l c c}
    \hline\hline
        Parameter & Symbol & Unit & Prior & Minimum & Maximum \\ \hline
        First/Second BH mass & $m_{1, 2}$ & $M_\odot$ & uniform & $10$ & $80$\\
        Spin vectors & $\boldsymbol{S}_{1,2}$ & $-$ & $-$ & $-$ & $-$ \\
        $\,-$ magnitudes &  & $-$ & uniform & $0$ & $0.99$\\
        $\,-$ polar angles &  & rad & \multirow{2}{*}{\Big{\}}\,uniform solid}  & $0$ & $\pi$ \\
        $\,-$ azimuthal angles &  & rad & & $0$ & $2\pi$ \\
        Luminosity distance & $d_L$  & Mpc & uniform radius & 10 & 1000 \\
        Right ascension & $\alpha$ & rad & \multirow{2}{*}{\Big{\}}\,uniform sky} & $0$ & $2\pi$  \\
        Declination & $\delta$ & rad &  & $-\pi/2$ & $\pi/2 $\\
        Binary inclination angle & $\theta_\text{JN}$ & rad & sin & $0$ & $\pi$ \\
        Coalescence phase angle & $\Phi$ & rad & uniform & $0$ & $2\pi$ \\
        Polarization angle & $\Psi$ & rad & uniform & $0$ & $2\pi$ \\
        Time shift & $t_\text{shift}$ & s & uniform & $-0.1$ & $0.1$ \\
    \hline\hline
    \end{tabular}
    \caption{Prior distribution $p(\bstattheta)$ on the signal model parameters.  }
    \label{tab:prior}
\end{table*}

\newpage
\section{Mock gallery}

\begin{figure}[h!]
    \hspace{-0.75cm}\includegraphics[width=1.1\textwidth]{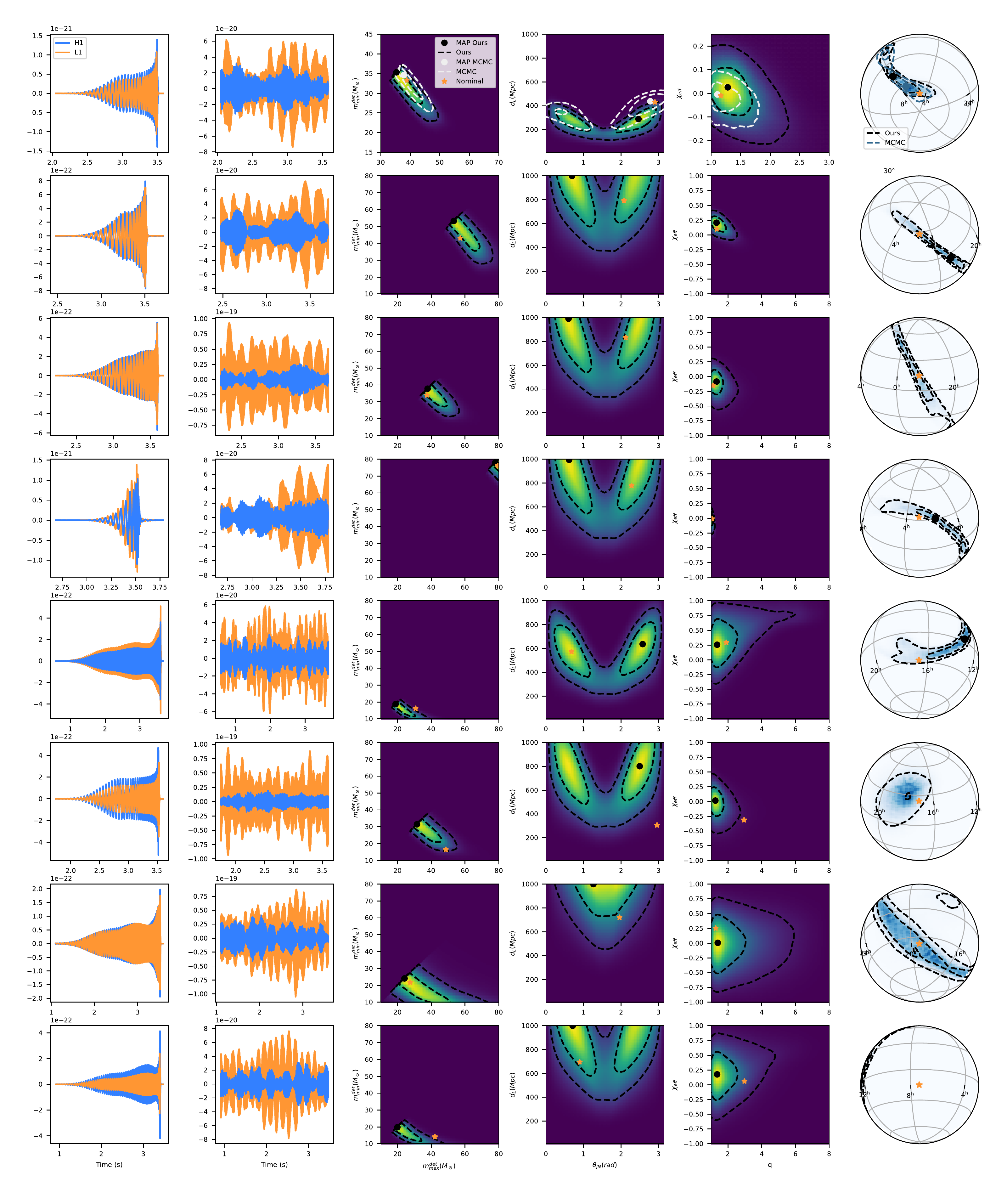}
    \caption{Results obtained with our method on simulated gravitational waves. The first column shows the gravitational wave without noise. The second column shows the same gravitational wave with noise added. 
    The last 4 columns show the results obtained when performing posterior parameter inference from the noisy gravitational wave. For each sample, the 50\% and 90\% credible intervals as well as the Maximum a posteriori estimates are shown. Those are compared to the nominal parameter values used to generate the data. 
    In the first row, our results are compared to results obtained using MCMC.}
    \label{fig:posteriors}
\end{figure}

\end{document}